\documentstyle[11pt,paspconf,psfig,epsf]{article}

\newcommand{\gtap}{\mathrel{\hbox{\rlap{\lower.55ex \hbox {$\sim$}}
                   \kern-.3em \raise.4ex \hbox{$>$}}}}
\newcommand{\ltap}{\mathrel{\hbox{\rlap{\lower.55ex \hbox {$\sim$}}
                   \kern-.3em \raise.4ex \hbox{$<$}}}}
\newcommand{\mdot}{\dot{M}}
\newcommand{\msun}{\, M_{\odot}}
\newcommand{\rsun}{\, R_{\odot}}

\begin{document}

\title{Observations of accretion disks in X-ray binaries}

\author{Frank Verbunt}
\affil{Astronomical Institute, Utrecht University, Postbox 80.000,
   3507 TA Utrecht, the Netherlands; email verbunt@phys.uu.nl}

\begin{abstract}
A brief overview is given of the observations of accretion disks in 
X-ray binaries, discussing their geometry,
their effect on the rotation period of the accreting object, 
their optical and X-ray spectra, time variability,
and radio emission and jets. Some outstanding questions for disk theory
are listed.
\end{abstract}


\section{Introduction}

The bright X-ray sources in the sky are located around the galactic plane.
X-ray and optical studies show that most of them are binaries in which
a neutron star or a black hole accretes matter from a companion.
Our knowledge of X-ray binaries is reviewed in Lewin et al. (1995b),
Tanaka \&\ Shibazaki (1996) and Chen et al.\ (1997).

In the high-mass X-ray binaries the mass donor is an O or B star with mass 
$M\gtap10\msun$.
Since high-mass stars have short life times, these binaries must be young,
$\ltap 10^7\,$yr.
The neutron star in such a binary has a strong magnetic field, 
$B\simeq 10^{12}\,$G, which funnels the accretion stream towards the magnetic 
poles, and causes these sources to show X-ray pulses with the rotation period
of the neutron star.
In high-mass X-ray binaries with short orbital  periods, $P_b\ltap 10\,$d,
the accretion onto the magnetosphere is thought to occur via an accretion
disk; in wider systems accretion may occur directly from the stellar wind
of the Be donor onto the magnetosphere.

In low-mass X-ray binaries the mass donor is a star with mass $M\ltap1\msun$.
Such binaries can be very old, $\gtap 10^9\,$yr.
Accretion generally occurs via a disk right onto a neutron star with a low
magnetic field, as derived from the absence of X-ray pulses, or onto a
black hole.
Quite a few low-mass X-ray binaries are soft X-ray transients, i.e.\ they
appear suddenly as bright X-ray sources, and then gradually disappear
again on a time scale of weeks to months.
The compact star in many transients is thought to be a black hole, but
some transients are known to contain a neutron star.
Neutron stars in low-mass X-ray binaries can be identified as such when 
an X-ray burst is observed, a flash of X-rays that lasts tens of seconds.

This article reviews observations of accretion disks in X-ray binaries.
The geometry of disks is discussed in Section 2, X-ray spectra in
Section 3, variability in Section 4, and radio emission and jets
in Section 5.
Remaining questions for theory are discussed in Section 6.
This review is far from complete, but tries to indicate how observations
help us study accretion disks.

\section{Geometry of accretion disks}

\subsection{High-mass systems}

A star in a close binary is deformed from a spherical shape, due to the orbital
revolution and to the gravitational attraction of its companion.
Equipotential surfaces inside each star are further apart on the side facing
the other star; as a result that side has a lower effective temperature.
The socalled ellipsoidal lightcurve of a binary star therefore shows
two maxima and two minima in each orbit, as the elongated and
foreshortened sides are observed in turn.
The deeper minimum occurs when we observe the side facing the companion.
If the companion is a very bright source, however, the surface of the
star facing it will be heated, and the corresponding minimum is filled
in. In extreme cases the heating dominates, and the orbital lightcurve
shows a single maximum that occurs when we observe the heated side.
The combination of ellipsoidal variation and heating cannot explain the 
observed orbital lightcurve of some X-ray binaries, like SMC~X-1 (Figure 1).

\begin{figure}[]
\centerline{
     \begin{minipage}[b]{3in}
          \psfig{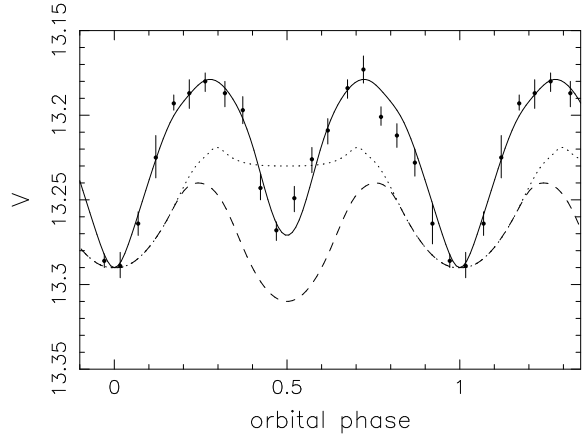}
     \end{minipage}
     \begin{minipage}[b]{2.4in}
          \caption{\hspace*{-1em}The observed orbital lightcurve
of SMC X-1 ($\bullet$) together with the predicted variations
due to ellipsoidal variation only ($--$), due to ellipsoidal
variation plus X-ray heating ($\cdots$), and due to these two
effects plus an accretion disk (---). 
After Tjemkes et al.\ (1986).}
     \end{minipage}
}
\end{figure}

The presence of an accretion disk affects the lightcurve in two ways:
the disk is a source of light by itself, and it shields part of the
surface of the companion from the X-ray heating.
In the case of SMC~X-1 a satisfactory explanation of the observed
optical lightcurve is obtained if one assumes  an accretion disk around the
neutron star, whose outer edge has a thickness
equal to some 20\%\ of the disk radius, $H_d\simeq0.2R_d$, and with
a surface temperature $T_d\simeq30000\,$K.
Only few high-mass X-ray binaries have been studied in sufficient
detail to estimate disk parameters (Tjemkes et al.\ 1986).

A strong magnetic field of a neutron star dominates the mass flow close to 
the neutron star.
A simple estimate of the radius $r_m$ of the magnetosphere is found by assuming
that the magnetic pressure at $r=r_m$ equals the ram pressure, 
$B^2/8\pi=\rho v^2$;
that the magnetic field is a dipole, $B(r)=B_o(R/r)^3$; and that the density
can be estimated from spherical inflow, $\rho=\mdot/(4\pi r^2v)$, at
freefall velocity $v=\sqrt{2GM/r}$. This gives
\begin{equation}
r_m = R  \left({B_o^2R^{5/2}\over 2\mdot\sqrt{2GM}}\right)^{2/7} 
\propto {B_o}^{4/7}{L_x}^{-2/7}
\end{equation}
where the latter proportionality uses $L_x=GM\mdot/R$.
In these equations $M$ and $R$ are the mass and radius of the neutron
star, $B_o$ is the magnetic field strength at the surface of the
neutron star, and $L_x$ is the X-ray luminosity due to an accretion
rate $\mdot$ onto the neutron star.
It can be argued that the magnetic field disrupts a disk at a
radius near $r_m$, and that the torque exerted by the accreting matter
on the neutron star is given by $N\simeq\mdot\sqrt{GMr_m}$.
For the rate of period  change this implies 
\begin{equation}
\dot P\propto N\propto \mdot^{6/7}\propto {L_x}^{6/7}
\end{equation}

The spinup of the neutron star stops when the rotation velocity
of the magnetosphere exceeds the Keplerian circular velocity at $r_m$, i.e.:
$\Omega r_m\equiv 2\pi r_m/P\geq\sqrt{GM/r_m}$, at which point matter 
is flung out rather than accreted.
The equilibrium period is found with the equal sign in the last equation:
\begin{equation}
P_{eq} =2\pi\left({{B_o}^2R^6\over 2\sqrt{2}\mdot}\right)^{3/7}
\left({1\over GM}\right)^{5/7}\propto {B_o}^{6/7}{L_x}^{-3/7}
\end{equation}
From these equations it was argued in the 1970s that systems with short
($\sim 1\,$s), secularly decreasing pulse periods, like
SMC~X-1 and Cen~X-3, contain accretion disks, and are close to
the equilibrium period $P_{eq}$ according to Eq.~3, whereas
systems with long pulse periods with no obvious secular change,
like Vela~X-1, were thought to accrete directly from a wind.
This agreed with the observations of optical orbital lightcurves,
discussed above, which show a disk in SMC~X-1 and Cen~X-3, but not in
Vela~X-1.

This simple picture must now be modified following the much more detailed
observations of spin evolution of some X-ray pulsars obtained with
BATSE on the Compton Gamma Ray Observatory (Bildsten et al.\ 1997).
These observations show that the slow long-term spinup of a system like
Cen~X-3 is composed of periods of much more rapid spinup alternated by
periods of almost equally fast spindown.
Furthermore, systems thought to accrete from a stellar wind occasionally
show fast spin changes.
And during the decline of transients the derivative of the pulse period
scales with
X-ray luminosity as $\dot P\propto L_x^n$, with $0.9<n<1.2$.
The order of magnitude of the spinup and spindown torques is as
estimated from Eq.~2. However, the changeovers from spinup to spindown
at intervals varying from 10-100 days in Cen~X-3 to 20 years in
4U$\,1626$-$67$ and GX$\,1$+$4$ are not understood.
The short epochs of rapid spinup of the wind fed system
GX$\,301$-$2$ are interpreted as due to the formation of a 
temporary accretion disk (Bildsten et al.\ 1997).

\subsection{Low-mass X-ray binaries}

The capability of the EXOSAT satellite to observe X-ray sources without
interruption for a period up to 4 days led to the discovery of orbital 
variations in the X-ray flux from several low-mass X-ray binaries
(Figure 2).
White \&\ Holt (1982) show that the X-ray lightcurve of 4U$\,1822$-$371$
can be explained by the obscuration of an extended X-ray source centered on 
the neutron star by an accretion disk which varies in height around its
outer edge.
A partial eclipse of the extended X-ray source by the companion star
is also observed.
The disk has a radius $R_d\simeq0.6\rsun$ (about 1/3 the distance
between the two stars) and a maximum thickness at the outer edge
$H_{max}\simeq0.15R_d$.
The orbital X-ray lightcurve of EXO$\,0748$-$676$ shows a complete
eclipse of the X-ray source by the companion star and discrete
absorption events, `dips', in addition to smooth variation.
The depth and location of the dips vary between orbits,
which indicates a disk thickness that varies in time.
4U$\,1820$-$30$, an 11$\,$min binary in the globular
cluster NGC$\,$6624, shows long-term variations in its orbital
X-ray lightcurve, and therefore presumably in the structure of the
outer rim of the accretion disk (Van der Klis et al.\ 1993).

\begin{figure}[]
\centerline{\psfig{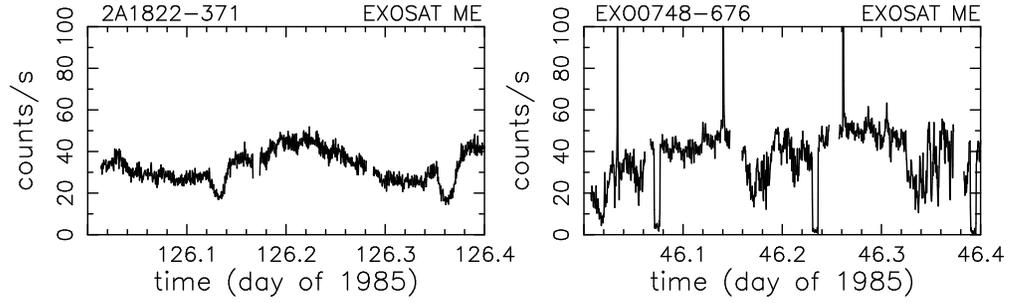}}
\caption{Orbital X-ray lightcurves of two low-mass X-ray binaries
  as observed with the medium-energy detector (1-10 keV) of EXOSAT. The smooth
  variation of the flux of 2A$\,1822$-$371$ (Hellier \&\ Mason 1989)
  is due to variable occultation
  of an extended source by the outer edge of the accretion disk; the
  events near day 126.13 and 126.36 are occultations of the same X-ray
  source by the companion star. The flux of EXO$\,0748$-$676$ 
  shows modulation caused by a disk of variable thickness, three complete
  eclipses by the companion star, and three X-ray bursts (which exceed the
  scale; Parmar et al.\ 1986).
\vspace*{0.2cm}}
\end{figure}

4U$\,1822$-$371$ has also been studied in the optical and infrared.
The lightcur- ves at these wavelengths can be modelled with a flat accretion
disk surrounded by a vertical rim which radiates from the inside and
(about 4 times less) from the outside, combined with a heated face
of the companion star. The disk radius and rim height agree
with the values found from the X-ray analysis (Mason \&\ C\'ordova 1982).

The X-rays from near the neutron star do not only heat the companion
but also the surface of the outer regions of the accretion disk.
The X-rays impinging on these regions are partially reflected, and
partially absorbed after which the energy is re-emitted at optical and 
ultraviolet wavelengths.
The effective temperature $T_e$ of the emission is set by the absorbed
energy: $\sigma {T_e}^4=(1-\epsilon)(L_x/4\pi r^2)\cos\xi$ where $\epsilon$
is the X-ray albedo, $L_x$ the X-ray luminosity and $\xi$ the angle between
the normal to the disk surface and the line connecting the disk surface
to the X-ray source.
In the observed range of temperatures, $T_e\sim 3\times10^4\,$K,
the emitted flux in the V band (5500\AA) scales as $S_V\propto {T_e}^2$,
hence $S_V\propto {L_x}^{1/2}/r$. As the total surface of the disk
scales with $r^2$, the total visual flux of the disk scales
as $L_V\propto {L_x}^{1/2}r\propto  {L_x}^{1/2}P_b^{2/3}$, where the
last proportionality uses Kepler's third law (and ignores the small
dependence on the masses of the stars). This relation gives a fair
description of the observed absolute visual magnitudes of low-mass 
X-ray binaries
as a function of their X-ray luminosities and orbital periods, which
indicates that the visual emission of these binaries is indeed dominated
by reprocession of X-rays (Van Paradijs \&\ McClintock 1994).

An X-ray burst on the neutron star heats the inner regions of the
disk first, and the outer regions later, due to the difference
in light travel time. As a result the optical flash following an X-ray
burst is both delayed and broadened with respect to the X-ray burst.
Bunk (1992) has shown that a disk with radius $0.64\rsun$ and
height $H(r)\propto r^{9/7}$ can reproduce the optical burst 
observed to follow an X-ray burst in 4U$\,1636$-$53$ (see also 
Matsuoka et al.\ 1984), provided reprocessing of the burst
by the companion star is also taken into account (see Figure 3).

\begin{figure}[]
\centerline{
     \begin{minipage}[b]{3in}
          \psfig{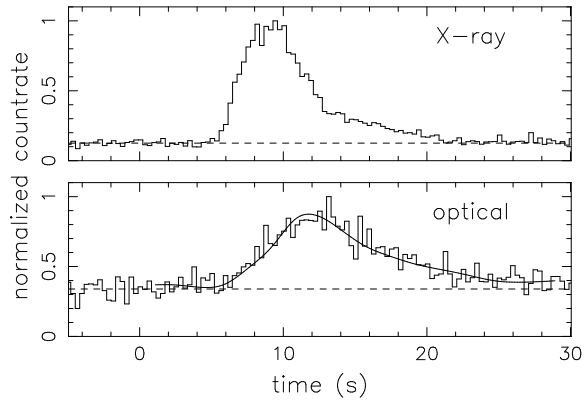}
     \end{minipage}
     \begin{minipage}[b]{2.4in}
          \caption{\hspace*{-1em}Lightcurve of an X-ray burst in 
  4U$\,1636$-$53$ and its reprocessing in the accretion disk observed
  by Tr\"umper et al.\ (1985). The dashed
  lines indicate the level before the burst, the smooth curve a model
  fit of reprocession in the disk {\it and} on the surface of the companion
  star. After Bunk (1992).}
     \end{minipage}
}
\end{figure}

When the X-ray emission of a soft X-ray transient has faded, the reprocessed
optical light has faded with it, and in several systems the mass donor
can then be detected.
In Cen~X-4 and in A$\,0620$-$00$ the spectrum of the donor can be
identified, with that of a K7$\,$V and K5$\,$V star respectively.
Subtraction of the stellar spectrum from the observed spectrum
shows a flat spectrum with Balmer emission lines, which indicates
that the accretion disk is still present and active when the 
transient X-ray source is in the low state (Figure 4).
In both systems the disk luminosity indicates a mass flow through the
outer disk $\mdot_{opt}\sim10^{-10}\msun$/yr (Chevalier et al.\ 1989,
McClintock et al.\ 1995). 
This is in marked contrast with the very low X-ray luminosity in the low
state, $10^{33}\,$erg/s in Cen~X-4 and $6\times 10^{30}\,$ergs/s in
A$\,0620$-$00$, which translate into accretion rates onto the central
objects of $\sim 10^{-13}\msun$/yr and $<6\times 10^{-15}\msun$/yr
respectively (Van Paradijs et al.\ 1987, McClintock et al. 1995).

\begin{figure}[]
\centerline{
     \begin{minipage}[b]{3in}
          \psfig{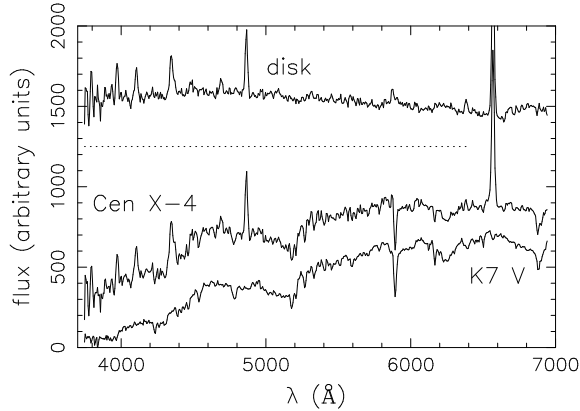}
     \end{minipage}
     \begin{minipage}[b]{2.4in}
          \caption{\hspace*{-1em}Quiescent optical spectra of
 Cen X-4, together with that of a K7V star. The difference
 between the two, also shown, is the spectrum of the quiescent 
 accretion disk. For clarity the disk spectrum is shifted,
 its zero level indicated with the dotted line. After Chevalier
 et al.\ (1989).}
     \end{minipage}
}
\end{figure}

\subsection{A special case: Her~X-1}

Her~X-1 is an X-ray pulsar in an orbit of $P_b=1.7\,$d with a donor of
$2\msun$, and thus neither a high-mass nor a low-mass system.  Its
optical lightcurve is dominated by the heated face of the donor, with
an eclipse of the accretion disk at minimum.  The X-rays show a period
of 35$\,$d, during ten of which the X-rays are on, followed by 25 days
of off state.  Interestingly, the donor is heated -- as shown by
the optical lightcurve -- also in the X-ray off state.  Detailed
analysis of complicated but regular variations in the optical
lightcurve show that these can be explained with an accretion disk
around the pulsar which is tilted 30$^{\circ}$ with respect to the
orbital plane and which precesses with a 35$\,$d period. The disk has
a radius $R_d\simeq2.3\rsun$ (a third of the distance between the two
stars), and a height at the outer edge $H_d\simeq0.3R_d$.
The off state of the X-rays occurs when the disk obscures the pulsar
for us, but not for the companion star (Gerend \&\ Boynton 1976).

\section{X-ray spectra}

The X-ray spectrum of an X-ray pulsar arises from the gas accreting
along the magnetic field lines close to the magnetic poles, and is 
characterized by broadband emission with small features, like K-shell
Fe emission near 7$\,$keV and cyclotron lines at higher energies.
The spectrum thus holds little information about the accretion disk,
which is disrupted at larger distances from the neutron star (Eq.$\,$1).

X-ray spectra of systems with disks that approach the compact
object more closely have been reviewed by Tanaka (1997). The following
discussion owes much to that review.
The satellites that provides the data for such studies are GINGA and
EXOSAT.
At high luminosities ($L_x\gtap10^{37}\,$erg/s) the X-ray
spectra of black hole binaries may be described with a sum of
black body spectra at different temperatures. 
These different temperatures are interpreted as corresponding to different 
radii in an accretion disk, where according to simple theory
$T_{eff}\propto r^{-3/4}$. (Note that irradiation of the disk may
be ignored close to the compact star; and that a correction term
related to the inner boundary condition is ignored here. Note also
that one doesn't expect the disk surface to emit black-body-like
spectra in the presence of strong electron scattering, as already 
argued by Shakura \&\ Sunyaev 1973.)
Typical inner disk radii $r_{in}$ derived by fitting this model to the
observed X-ray spectra are $\sim 50\,$km, of the order of
the innermost stable orbit around a black hole.
Remarkably, $r_{in}$ doesn't change when the X-ray luminosity of the
black hole system changes by up to four orders of magnitude.

X-ray spectra of neutron star systems have the same multi-temperature disk
component, but with an inner radius $r_{in}\sim10\,$km, or the order
of the radius of the neutron star.
In addition these sources show a soft black body spectrum modified by
electron scattering, with a colour temperature $kT_c\sim2.4\,$keV.
This soft component is assigned to the neutron star surface.

When the X-ray luminosity drops below $\sim 10^{37}\,$erg/s a power-law 
component appears in the spectra both of black hole and of neutron star
systems.
At lower luminosities the photon number spectrum is dominated by the power law,
$N(E)\propto E^{-n}$, with $n\simeq 1.6$ for black hole systems and
$n\simeq1.8$ for neutron star systems.
The transition from a multi-temperature spectrum to a power law
is thought to occur when the accretion disk changes from an optically
thick gas to a hot optically thin plasma. The thin plasma spreads over
a larger volume.
Hence, a burster whose burst spectrum corresponds to a pure
black-body atmosphere
at high $L_x$, shows burst spectra with a significant Comptonization tail
at lower $L_x$ when the persistent emission is dominated by the power-law
spectrum (Nakamura et al.\ 1989). 

Direct evidence of the spatial extent of the power-law component has been found
from the analysis of the specta during the dips in the orbital X-ray
lightcurves (Sect.\ 2.2). The variation of the spectrum of EXO$\,0748$-$676$
observed with ASCA can be described as due to variable absorption
of the point-like black-body component and of part of the 
power-law component; part of the power law component remains unabsorbed.
The variable absorption is presumably due to matter near the disk edge;
the 2.2$\,$keV black body to the neutron star; and the power law component,
with index $n=1.7$,
must be sufficiently extended that part of it escapes eclipse.
A Gaussian line at 0.65$\,$keV is affected by the same partial absorption
as the power-law component (Church et al.\ 1998).

In low-mass X-ray binaries with a neutron star,
subtle differences in spectral changes as a function of luminosity
were found  by Hasinger \&\ Van der Klis (1989) to be correlated with 
differences in (the Fourier transforms describing) the rapid variability,
and led these authors to discriminate Z- and Atoll sources. They also 
hypothesized that the difference between these two classes is due to
a slightly higher magnetic field of the neutron stars in the Z-sources.
However, no direct evidence for a magnetic field has been found in any of
these systems.
The relation between the spectral changes that discriminate Z- and
Atoll sources to the spectral changes as a function of luminosity
reviewed by Tanaka (1997) -- who does not discriminate between
these two classes -- deserves closer enquiry.
White et al.\ (1988) discuss X-ray spectra observed with EXOSAT,
and criticize the models based on optically thick multi-temperature
disks; they show that Comptonization of soft photons gives a more consistent
description of the observations, also at high luminosities.

As the X-ray luminosity drops below $\sim 10^{36}\,$erg/s further spectral 
changes take place. ROSAT observations show that the soft part of the 
spectrum of both neutron-star and black-hole systems can be described with
a black body spectrum (e.g.\ for Aql~X-1, Verbunt et al.\ 1994; for
A$\,0620$-$00$, McClintock et al.\ 1995).
The black body describes the X-ray spectrum in a remarkable range
of luminosities, from $10^{35}$ to $10^{32}\,$erg/s in Aql~X-1.
The associated emission area is small, $\sim\,$km$^2$, at the lowest
luminosities.
ASCA observations of Cen~X-4 show that the spectrum at energies above
$\sim 2\,$keV is dominated by a power law (Asai et al.\ 1996).
The ASCA spectrum also shows that the black body component does not show
strong lines, i.e.\ that it arises in optically thick emission
(Tanaka 1997).

The spectral fits discussed above all use rather simple spectral forms as 
black bodies and power laws. High-quality spectra of the brightest sources 
show spectral features that belie these simple descriptions. Clear evidence
for spectral lines and edges have been found e.g.\  with the Einstein
satellite for Sco~X-1
and with EXOSAT for Cyg~X-2 (Vrtilek et al.\ 1991a, Chiappetti et al.\ 1990).
Hopefully, high-quality spectra obtained with AXAF and XMM will allow
more sophisticated modelling.

\section{Variability}

X-ray binaries are variable on many time scales.
The orbital variability reflects the changing aspect of the binary as
it revolves.
The variability discussed in this section is intrinsic, and refers to
real changes in the binary.

It has been mentioned above that many low-mass X-ray binaries are
transients, appearing as bright X-ray sources on a time scale of days and
gradually becoming fainter again on a time scale of weeks to months.
The variation in X-ray luminosity is accompanied by variation in the
X-ray spectrum (Sect.\ 3).
It has been argued that the flux decline accelerates at
$L_x\ltap 10^{36}\,$erg/s in Aql~X-1, and that this is due to prohibition
of accretion by centrifugal forces exerted by the magnetic field of the 
neutron star (Campana et al.\ 1998).
The problem with this explanation is that black hole systems show
rather similar lightcurves (e.g.\ GRS$\,$J$\,$1655$-$40, M{\'e}ndez et 
al.\ 1998).
Kuulkers et al.\ (1996) show that the outburst lightcurve of 
A$\,0620$-$00$ is 
remarkably similar to that of a class of dwarf novae with very strong
outbursts, e.g.\ in showing a rapid drop at the end of the outburst,
followed by several less bright and shorter outbursts.

In the low state, variability of the flux by factors $\sim 3$ on
a a time scale of hours has been detected both in the optical
(e.g.\ Cen~X-4, Chevalier et al.\ 1989) and in X-rays
(Campana et al.\ 1997), indicating a variation of the mass flow
onto both in the outer disk and in the inner disk. (Note that irradiation
of the disk is not important at these low X-ray luminosities.)
The average interval between outbursts of Aql~X-1 varied from
$\sim125\,$d in the 1970s, as observed with the Vela satellites,
to $\sim310\,$d in the 1980s, as observed with GINGA (Kitamoto et 
al.\ 1993).

\begin{figure}[]
\centerline{\psfig{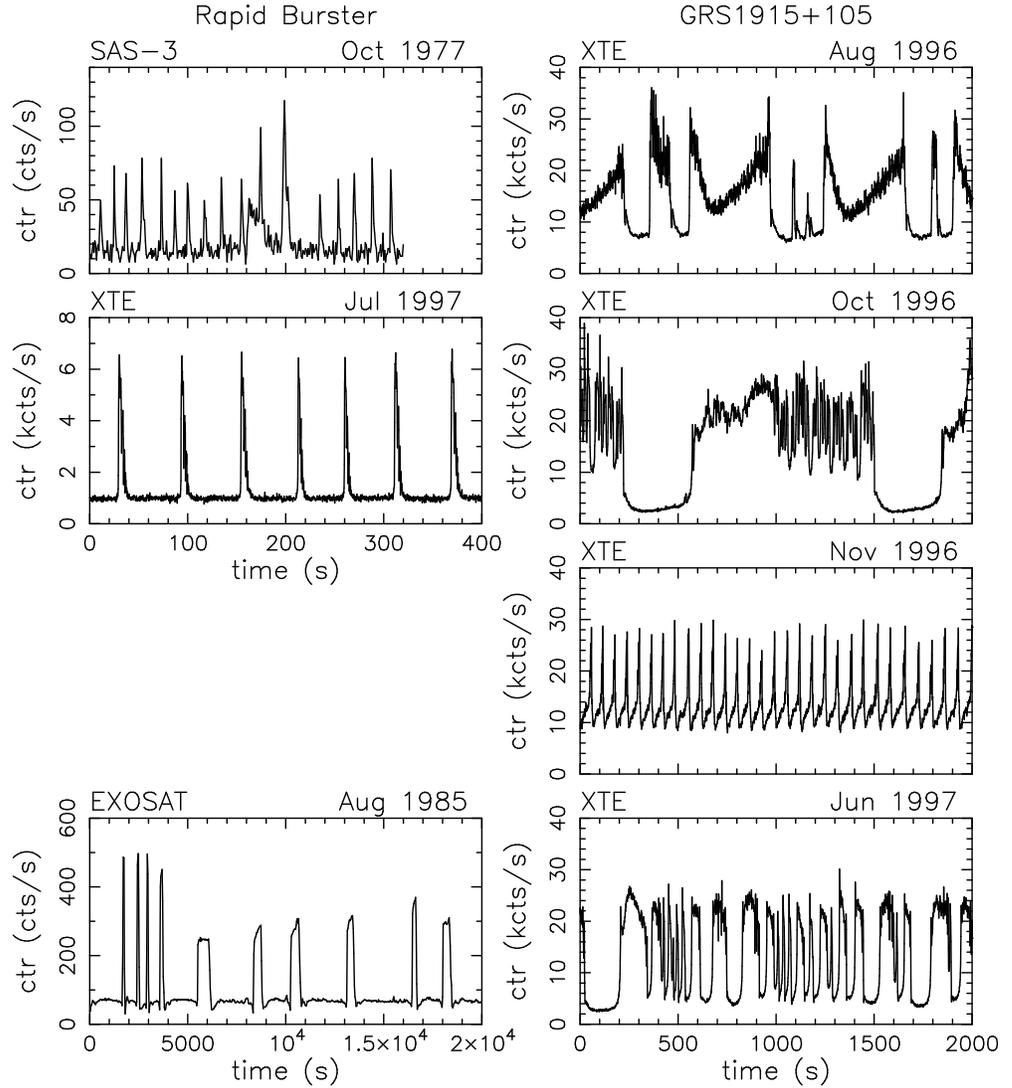}}
\caption{Long term variations in the X-ray lightcurves of
the Rapid Burster (left, top two frames on same time scale) and of
GRS$\,1915$+$105$ (right, all frames on same scale).
A thermonuclear burst shows up between the rapid bursts of Oct 1977
at $t=160\,$s. After Hoffman et al. (1977), Guerriero et al.\ (1998),
Belloni et al.\ (1997).}
\end{figure}

Variability on a very short time scale is displayed by the 'Rapid Burster',
a recurrent transient X-ray source in the globular cluster Liller~1.
When active, this source emits its X-rays in discrete events, bursts. 
These bursts were fast and frequent when the source was discovered,
giving it its name; at other times the bursts were much longer and
less frequent (Figure 5). The source has also been detected
in a state of persistent emission. The rapid bursts are interpreted
as discrete accretion events; that the accreting star is a neutron star
is shown by occasional thermonuclear bursts. For a review of this remarkable
source, see Lewin et al.\ (1995a).
A stupendous range of variability is also displayed by the transient
GRS$\,1915$+$105$ in observations obtained with XTE (Figure 5).
Multi-temperature disk models are fitted to the X-ray spectrum of the
June 1997 observations by Belloni et al.\ (1997) who derive an inner
disk radius $r_{in}\sim20\,$km for the high state, and $r_{in}\sim80\,$km
for the low state.
Again, it should be cautioned that the physical meaning of multi-temperature
black-body spectra is unclear (Section 3; White et al.\ 1988).

Sensitive study of variability in low-mass X-ray binaries with use of 
Fourier transforms discovered white noise, red noise and quasi-periodic
oscilations, related to subtle changes in the X-ray spectra.
With XTE quasi-periodic oscillations were discovered at kHz frequencies, 
occasionally at two frequencies simultaneously.
The most commonly considered explanation for quasi-periodic oscillations
considers them to be the result of a beat between the rotation
frequency of the neutron star an the revolution frequency of a gas
blob in the accretion disk close to the magnetospheric radius.
This model predicts a constant distance,
corresponding to the frequency of the rotation of the neutron star,
between the frequencies of the two high-frequency oscillations.
The observation that the difference is in fact variable has thrown this
theoretical model into some disarray (Psaltis et al\ 1998).

\section{Radio emission and jets}

Radio emission by low-mass X-ray binaries is quite common, as reviewed
by Hjellming \&\ Han (1995).
A spectacular case is SS$\,$433, which emits two opposite jets with 
velocities $v\simeq0.26c$ which precess with a period of 164$\,$d.
The jets are also detected in the optical (where they were first discovered)
and in X-rays. Their precession is deemed to arise from the precession
of the accretion disk from which they are launched. SS$\,$433 is reviewed
by Margon (1984). 

Two radio jets emitted by GRS$\,1915$+$105$ expand superluminally with an 
intrinsic flow velocity of 0.92$c$ (Mirabel \&\ Rodriguez 1994).
A 2.2$\,\mu$m flare followed $\sim 15\,$min later by a radio flare was
observed to start during the low flux level of the repeating X-ray variability
pattern in September 1997 (Mirabel et al.\ 1998).
It would be tempting to conclude that the accretion energy
is spent in the emission of a hot expanding plasma cloud rather than
in X-rays at the the low X-ray flux level; perhaps one should await
more instances of correlated X-ray, $\mu$m and radio emission
before drawing definite conclusions.

\section{Conclusions}

It may be clear from the above that there is no dearth of observations
of accretion disks in X-ray binaries, albeit that most of them are 
tantalizingly indirect, thus giving the theorists a longer leash than is
perhaps good for them.
An important guide in judging theoretical explanations is the discrimination
of properties common to many systems and those of only some systems,
occasionally even of one unique system.
Thus, explanations of properties of black hole systems that
make essential use of the black hole presence are suspect if the same
properties are also observed in neutron star systems.
Conversely, an explanation that does not use parameters unique to a system
like the Rapid Burster, SS$\,$433 or GRS$\,1915+105$ cannot be used to
explain the unique properties of such a system.

I will end this review by mentioning some of the more pressing questions
raised by the observations of accretion disks in X-ray binaries.

Why is the accretion disk thicker in a low-mass X-ray binary than in a
cataclysmic variable? Is this related to the mass flow through the
disk, as suggested by variations in disk thickness in the dwarf nova OY~Car
(Marsh, these proceedings) or to the effect of irradiation
(Vrtilek et al.\ 1991b)?
What causes the disk to precess in Her~X-1 and SS$\,$433 and several
other systems (Pringle, Larwood, these proceedings)?
How can the interaction between a disk and the magnetic field of a
neutron star lead to alternating periods of spinup and spindown of
very similar magnitude, switching on times scale ranging from weeks
in some systems to decades in others (Pringle, Yi, these proceedings).

What is the nature of the X-ray spectra and how does the spatial extent of the
power law component arise? Why do the spectra change with X-ray 
luminosity (Yi, these proceedings)?
Why does the average interval between outbursts vary in one system
(125 to 309 days in Aql~X-1) and how can it range from such short intervals 
for Aql~X-1 to $\sim50\,$yr in other transients (Livio, these
proceedings)?
The semi-regular variability of the Rapid Burster and of
GRS$\,1915+105$ as well as quasi-periodic oscillations in many
low-mass X-ray binaries have not been convincingly explained.
Are these variations related to the emission of radio jets (Blandford, these
proceedings)?

\acknowledgments 

I am grateful to Wolfram Bunk, Claude Chevalier, Tomaso Belloni, Walter Lewin
and Derek Fox for sending me data used in Figures 3, 4 and 5.


\end{document}